\documentstyle[preprint,aps]{revtex}

\begin{document}
\tightenlines
\draft
\title{Temperature dependence of electron transport \\
through a quantum shuttle}
\author{Anatoly Yu. Smirnov}
\address{D-Wave Systems Inc., 320-1985 W. Broadway \\
Vancouver, British Columbia Canada V6J 4Y3}
\author{Lev G. Mourokh}
\address{Department of Physics and Engineering Physics, \\
Stevens Institute of Technology, Hoboken, NJ 07030 }
\date{\today }
\maketitle

\begin{abstract}
{We analyze electron transport through a quantum shuttle for the applied voltage below the instability threshold. We obtain current-voltage characteristics of this system and show that at low temperature they exhibit pronounced steps. The temperature dependence of the current is calculated in the range from 2K to 300K and it demonstrates a wide variety of behavior - from $1/T$ decreasing to an exponential growth - depending on how deep the shuttle is in quantum regime. The results obtained are compared to experimental data on electron transport through long molecules.
}
\end{abstract}


Recent achievements in nanotechnology facilitate development of a new generation of nanodevices incorporating mechanical degrees of freedom, nanoelectromechanical systems (NEMS) \cite{Roukes1,Cleland1,Irish1}. In this, one of the most promising device structure is a single-electron shuttle \cite{Gorelik1} carrying an electron flow between two contacts. Even classical aspects of this model were shown to be important for a description of electron transport through a nanomechanical bell \cite{Blick1} and a $C_{60}$ single-molecule transistor \cite{Park1}. Theoretical analysis of such a system in the classical framework was done in Refs. \cite{Fedorets1,Nishiguchi1,Armour2}. However, many properties of NEMS can only be understood on a fully quantum mechanical basis. Quantum regime of the shuttle having mass $m$ and characteristic frequency $\omega_0$ occurs when an amplitude of zero-point mechanical fluctuations, $\Delta x_{zp} = \sqrt{\hbar /2 m \omega},$ is of order
of a tunneling length $\lambda$ involved in dependence of the tunnel
matrix elements on the distance $x$ between the shuttle and the leads, $T_{k\alpha}(x) = T_{k\alpha} \exp(x/\lambda_{\alpha})$ ($\alpha = L,R; \lambda_L =
-\lambda, \lambda_R = \lambda $). Effects of quantization of the mechanical motion were examined in Refs. \cite{Boese1,Armour1,McCarthy1} and the occurrence of the shuttle instability was predicted in Refs. \cite{Novotny1,Fedorets2} for the zero temperature case. 

In the present paper we examine the electron transport through the shuttle system at low applied bias voltage (below the instability threshold) with the special stress on the temperature effects. To accomplish this, we perform a combined theoretical analysis of mechanical motion and electron transport in the model of the quantum shuttle carrying a single electron level (strong Coulomb blockade regime). We show that at low temperature current-voltage characteristics exhibit pronounced steps. The first step takes place when the applied voltage compensates the initial separation between the single electron energy level in the shuttle. The second step occurs when the applied voltage facilitates electron tunneling from the left lead to the shuttle with absorption of virtual "quantum" of mechanical motion (phonon) by electron with consequent emission of this "phonon" during electron tunneling from the shuttle to the right lead. Moreover, we demonstrate that there is instability of electron transport through the shuttle, when the applied voltage is enough to produce this "phonon" absorbed by the shuttle system.

We show that the character of the current through the system as a function of temperature is strongly dependent on quantum parameter 
$$\nu_0 = {\hbar
\over 2 m\omega_0\lambda^2}$$ 
which defines a relative level of zero-point
mechanical fluctuations. In particular, we demonstrate that systems with
relatively small zero-point mechanical fluctuations $(\nu_0 \leq 0.1)$
exhibit weak temperature dependence of the current-voltage characteristics $%
I(V,T)$, whereas for the quantum shuttle with $\nu_0 \geq 0.4$ we obtain the current as the exponential function of temperature. It should
be emphasized that the model of the quantum shuttle can be of prime importance for theoretical
explanation of some aspects of electron transport through single molecules and
self-assembled monolayers \cite{Ratner1}. An organic molecule is coupled to the
leads with elastic links \cite{Reed1} and may oscillate as a whole. These
oscillations are reflected in the temperature dependence of the
current-voltage characteristics of the system \cite{ourMech}. It should be
noted that experimental current-voltage characterizations of organic
monolayers demonstrate a wide variety of temperature behaviour - from a
weak temperature dependence for molecules C8, C12, and C16 sandwiched
between Au contacts \cite{Reed2} to an exponential temperature dependence of
a conductance and current that has been observed by D. Stewart {\it et al} 
\cite{Stewart1} in experiments with a Langmuir-Blodget monolayer of eicosanoic acid (C20) connected to Pt electrodes. The latter results have not yet explained because only no temperature dependence (in the case of direct tunneling) or activation dependence (in the case of hopping transport) were expected. We suggest that the mechanical molecular motion may be the reason for such exponential dependence. The difference in measured temperature dependences probably reflects the difference in connections of the molecules to electrodes and, consequently, the different parameters $\nu_0$ for these systems.

To examine the electron transport through the quantum shuttle, we start from the Hamiltonian having the form ($\hbar =1, k_B = 1$)
\begin{eqnarray}
H = {\frac{p^2 }{2m}} + {\frac{m\omega_0^2 x^2 }{2}} + E_0 a^+a +  \nonumber
\\
\sum_{k \alpha } E_{k\alpha } c_{k\alpha }^+c_{k\alpha } - \sum_k (
T_{k\alpha} c_{k\alpha }^+a + T_{k\alpha}^* a^+c_{k\alpha } )
e^{x/\lambda_{\alpha}}.
\end{eqnarray}
Here $\alpha = L,R, \lambda_L = -\lambda, \lambda_R = \lambda ,$ and $V$ is a
voltage applied to the leads \cite{eVd}. The equations of motion, derived from the Hamiltonian, Eq.(1), are given by 
\begin{eqnarray}
i \dot{a} = E_0 a - \sum_{k\alpha} T_{k\alpha}^* c_{k\alpha }
e^{x/\lambda_{\alpha}} ,  \nonumber \\
i \dot{c}_{k\alpha } = E_{k\alpha } c_{k\alpha } - T_{k\alpha} a
e^{x/\lambda_{\alpha}},  \nonumber \\
\ddot{x} + \omega_0^2 x = \sum_{k\alpha }{\frac{1 }{m\lambda_{\alpha}}}(
T_{k\alpha} c_{k\alpha }^+a + T_{k\alpha}^* a^+c_{k\alpha } )
e^{x/\lambda_{\alpha}}.
\end{eqnarray}
It should be noted that we take into account nonlinearities associated with a nonlinear exponential dependence of the tunneling elements on the distance between the shuttle and the leads. 

The electric current through the shuttle is defined
as $I = I_L = - I_R$, where 
\begin{equation}
I_{\alpha } = e{\frac{d }{dt}} \sum_k \langle c_{k\alpha }^+c_{k\alpha
}\rangle = - i e \sum_k T_{k\alpha}^* \langle a^+ e^{x/\lambda_{\alpha}}
c_{k\alpha } \rangle + h.c.
\end{equation}
With the expression for the electron amplitude in the lead derived from Eq.(2), as 
\begin{equation}
c_{k\alpha }(t) = c_{k\alpha }^{(0)}(t) + i T_{k\alpha} \int dt_1
e^{-iE_{k\alpha }(t-t_1)} a(t_1) e^{x(t_1)/\lambda_{\alpha}}\theta(t-t_1),
\end{equation}
and using the formula 
\begin{eqnarray}
\langle a^+(t) e^{x(t)/\lambda_{\alpha}} c_{k\alpha }^{(0)}(t) \rangle = 
\nonumber \\
-iT_{k\alpha} \int dt_1 f_{\alpha} (E_{k\alpha})e^{-iE_{k\alpha }(t-t_1)}
\theta(t-t_1) \langle [a^+(t) e^{x(t)/\lambda_{\alpha}}, a(t_1)
e^{x(t_1)/\lambda_{\alpha}}]_+ \rangle,
\end{eqnarray}
resulted from the Langreth theorem, we obtain the electric current, as given by
\begin{eqnarray}
I_{\alpha } = - e\sum_k |T_{k\alpha}|^2 \int dt_1 f_{\alpha}
(E_{k\alpha})e^{-iE_{k\alpha }(t-t_1)} \theta(t-t_1) \langle [a^+(t)
e^{x(t)/\lambda_{\alpha}}, a(t_1) e^{x(t_1)/\lambda_{\alpha}}]_+ \rangle + 
\nonumber \\
\sum_k |T_{k\alpha}|^2 \int dt_1 e^{-iE_{k\alpha }(t-t_1)} \theta(t-t_1)
\langle a^+(t)a(t_1)\rangle \langle e^{x(t)/\lambda_{\alpha}},
e^{x(t_1)/\lambda_{\alpha}}\rangle + h.c.
\end{eqnarray}
Electrons in the $\alpha-$lead are assumed to be described by the Fermi
distribution $f_{\alpha }(E) = f(E-\mu_{\alpha }) = \left[ \exp\left( {\frac{%
E-\mu_{\alpha }}{T}}\right) + 1 \right]^{-1}. $ For averaged values of
electron correlators we have approximate expressions $\langle
a^+(t)a(t_1)\rangle = N e^{iE_0(t-t_1)}, \langle a(t_1)a^+(t)\rangle = (1-N)
e^{iE_0(t-t_1)} $ with $N = \langle a^+ a\rangle$ being a steady-state
electron population of the shuttle that can be found from the condition $%
I_R+ I_L =0.$ Using previously derived formulas (see Appendix in Ref.\cite
{ourMech}) in the wide-band limit we obtain the expression for the current, as
\begin{eqnarray}
I_{\alpha } = e^{\nu_c} e\Gamma_{\alpha } \sum_{m=-\infty}^{+\infty}
\sum_{l=-\infty}^{+\infty} J_m(\nu_0) I_l(\nu_c)\{ N[ 1 - f_{\alpha }(E_0 -
m\omega_0 - l\omega_0) ] -  \nonumber \\
(1-N)f_{\alpha }(E_0 + m\omega_0 + l\omega_0) \}.
\end{eqnarray}
Here, $J_m(\nu_0)$ and $I_l(\nu_c)$ are usual and modified Bessel functions
of order $m$ and $l$, respectively, $\Gamma_{\alpha } = 2\pi \sum_k
|T_{k\alpha}|^2 \delta (\omega - E_{k\alpha}),$ and $\nu_c = \langle \tilde{x}^2\rangle /\lambda^2$ is a
relative dispersion of mechanical oscillations for the nonequilibrium case. In the following, we
consider symmetric coupling between the shuttle and the leads, $\Gamma_L =
\Gamma_R = \Gamma.$ A steady-state value of the electron population in the
shuttle is determined by the relation 
\begin{equation}
N =C/D,
\end{equation}
with 
\begin{eqnarray}
C = \sum_{ml} J_m(\nu_0) I_l(\nu_c)[ f_{L}(E_0 + m\omega_0 + l\omega_0) +
f_{R}(E_0 + m\omega_0 + l\omega_0)]  \nonumber \\
\text{and} \nonumber \\
D = \sum_{ml} J_m(\nu_0) I_l(\nu_c) [ 2 - f_{L}(E_0 - m\omega_0 - l\omega_0)
+ f_{L}(E_0 + m\omega_0 + l\omega_0) -  \nonumber \\
f_{R}(E_0 - m\omega_0 - l\omega_0) + f_{R}(E_0 + m\omega_0 + l\omega_0)].
\end{eqnarray}
In the case of relatively small zero-point mechanical fluctuations, $\nu_0
\ll 1, \sum_l I_l(\nu_c ) = e^{\nu_c},$ we have the simple formulas for the
shuttle population $N$ and for the current $I$, as 
\begin{eqnarray}
N = e^{-\nu_c} \sum_l I_l(\nu_c) {\frac{ f_L(E_0 + l\omega_0) + f_R(E_0 +
l\omega_0) }{2}},  \nonumber \\
I = I_L = e\Gamma e^{\nu_c} \sum_l I_l(\nu_c) {\frac{ f_R(E_0 + l\omega_0) -
f_L(E_0 + l\omega_0) }{2}}.
\end{eqnarray}
It should be noted that
effective temperature of the {\it biased} nanomechanical system, which is
determined by the dispersion of mechanical fluctuations, $\langle \tilde{x}^2\rangle,$ can be different from the equilibrium temperature $T$ \cite
{ourMech}, so that we cannot neglect nonlinearities caused by $\nu_c$ even when $\nu_0 \ll 1$.

To calculate a relative dispersion of mechanical fluctuations, $\nu_c = \langle 
\tilde{x}^2 \rangle /\lambda^2 ,$ the equation (2) for the shuttle position
can be rewritten in the form of quantum Langevin equation 
\begin{equation}
\ddot{x} + \gamma \dot{x} + \omega_0^2 x = \xi,
\end{equation}
with a damping rate 
\begin{equation}
\gamma = \nu_0 \Gamma e^{\nu_c} B(\omega_0),
\end{equation}
 and a
fluctuation source, 
\begin{equation}
\xi = \sum_{ k \alpha }(T_{k \alpha }^*/ m
\lambda_{\alpha } ) \{ a^+ e^{x/\lambda_{\alpha}}, c_{k\alpha }^{(0)} \} +
h.c., 
\end{equation}
 which is characterized by the spectrum $K(\omega ) = \nu_0 e^{\nu_c}
\Gamma (\hbar \omega_0 / m) A(\omega_0).$ Here, we introduce the following
dimensionless combinations of the Fermi distributions, $A(\omega)$ and $%
B(\omega)$, as
\begin{eqnarray}
A (\omega ) = \sum_{\alpha }\sum_{ml} J_m(\nu_0)I_l(\nu_c) \times  \nonumber
\\
\{ N [ 2 - f_{\alpha}(E_0 - m\omega_0 - l\omega_0 - \omega ) -
f_{\alpha}(E_0 - m\omega_0 - l\omega_0 + \omega)] +  \nonumber \\
(1-N) [ f_{\alpha}(E_0 + m\omega_0 + l\omega_0 + \omega) + f_{\alpha}(E_0 +
m\omega_0 + l\omega_0 - \omega)] \}  \nonumber \\
\text{and} \nonumber \\
B(\omega ) =\sum_{\alpha} \sum_{ml} J_m(\nu_0)I_l(\nu_c) \times  \nonumber \\
\{ N [ f_{\alpha}(E_0 - m\omega_0 - l\omega_0 - \omega) - f_{\alpha}(E_0 -
m\omega_0 - l\omega_0 + \omega)] +  \nonumber \\
(1-N) [ f_{\alpha}(E_0 + m\omega_0 + l\omega_0 - \omega) - f_{\alpha}(E_0 +
m\omega_0 + l\omega_0 + \omega)] \}.
\end{eqnarray}
The relative level of mechanical fluctuations, $\nu_c$, can be found from
the self-consistent equation: 
\begin{equation}
\nu_c = \nu_0 { A(\omega_0)\over B(\omega_0)},
\end{equation}
supplemented by the equation (8) for the steady-state electron population $N$
of the shuttle.

We have solved self-consistent set of equations, Eqs. (8) and (15), and substituted the obtained values of $N$ and $\nu_c$ into Eq. (7) for electric current. Corresponding current-voltage characteristics are shown in Figure 1 for various temperatures, energetic separation between the electron level in the shuttle, $E_0$ and equilibrium chemical potential of the leads, $\mu$, being $0.2\omega_0$, $\omega_0/2\pi =1THz$, and for quantum parameters $\nu_0=0.7$ (Fig.1(a)) and $\nu_0=0.07$ (Fig.1(b)). It is evident from these Figures that there are pronounced steps in the low-temperature current-voltage characteristics at 
the bias voltages (i) $eV/2 = E_0 - \mu$ and (ii) $eV/2 = \omega_0 - E_0 + \mu$. The first one occurs when the chemical potential of the left lead passes through the energetic level of the shuttle, while the second one corresponds to the passing of the chemical potential of the right lead through the virtual level with energy $E_0 - \omega_0$. In the latter case, tunneling of an electron of the left lead having energy $E_0 - \omega_0$ to the shuttle level with energy $E_0$ is accompanied by absorption of virtual "quantum" of mechanical motion (phonon) with further tunneling of this electron to the state of the right lead having energy $E_0 - \omega_0$ accompanied by emission of this "phonon". This second step is much more pronounced for the case of larger $\nu_0$. When the chemical potential of the left lead passes through the virtual level with energy $E_0 + \omega_0$ (at the voltage $eV/2 = \omega_0 + E_0 - \mu$), electron tunneling from the left lead to the shuttle can be accompanied by absorption of such a phonon by the shuttle, and the system becomes unstable. To illustrate this, we present in the Inset of Fig.1(a) the current-voltage characteristics (in the logarithmic scale) with various values of $E_0 - \mu$. It should be noted that the condition for this instability is identical with that found in Ref. \cite{Fedorets2}. Finally, one can see in Fig.1(a) that current-voltage characteristics become more smoother and steps disappear with temperature increasing. 

The temperature dependence of the current through the quantum shuttle system is of special interest. It is shown in Figure 2 in the logarithmic scale for $E_0-\mu =0.2\omega_0$, $eV=0.4\omega_0$, $\omega_0/2\pi =1THz$, and various values of quantum parameter $\nu_0$. It is evident from this Figure that initial decrease of the current with temperature increasing is followed by the strong exponential growth at larger values of $\nu_0$. The exponential temperature dependence of the current-voltage
characteristics has been also predicted for the model of a tunnel junction
having its matrix elements modulated by the vibrational motion \cite{ourMech,ourSF}. 
In this model \cite{Mozyrsky}, the position of the
oscillator affects directly a probability of electron tunneling between
leads and the temperature dependence involved in distribution functions of
the leads is integrated out. Correspondingly, the exponential temperature
dependence, caused by fluctuations of an oscillator displacement, occurs for
any levels of zero-point mechanical fluctuations, and, in particular, for $%
\nu_0 \ll 1.$ The present study reveals that an inclusion of a
single-electron state into the vibrating molecule leads to an additional
factor $1/T$ in the formulas for current and conductance (caused by the temperature dependencies of the distribution function of electrons in the leads) that smooth out the exponential T-dependence in the case of small $\nu_0.$ Indeed, for high temperatures, $(T \gg |E - \mu_\alpha|,$ one can see that $\nu_c = \nu_0(2T/\hbar\omega_0)$, so that the current, Eq.(7) is
described by the formula 
\begin{equation}
I = e\Gamma \exp(4T\nu_0/\hbar \omega_0){eV\over 4 T},
\end{equation}
 where a temperature dependence reveals itself not only in the exponent $%
e^{4T\nu_0/\hbar \omega_0}$ as for the tunnel junction coupled to the mechanical
oscillator \cite{ourMech,ourSF}, but also in the factor $1/T$. An outcome of a competition between these two factors depends crucially on
the value of the relative level of zero-point mechanical fluctuations $\nu_0$. At small $\nu_0$ the factor $1/T$ dominates, whereas at larger $\nu_0$ the current-temperature curve demonstrates a very pronounced exponential temperature dependence. However, $\nu_0$ depends on the inverse tunneling length of system squared and even small variations of the tunneling length can bring about completely different behavior. Accordingly, the experimental data of Refs. \cite{Reed2,Stewart1} might correspond to slightly different conditions of connections between molecules and leads and still exhibit very different temperature dependence of the electron current.

In conclusion, we have analyzed electron transport through a quantum shuttle having single electron energetic level. We have shown that at low temperature there are pronounced steps in current-voltage characteristics corresponding to direct tunneling through the shuttle and to tunneling accompanied by consequent absorption and emission of virtual "quantum" of mechanical motion (phonon) by electron. We have found that when the applied voltage facilitates the emission of "real phonon" during electron tunneling from the left lead to the shuttle, the instability of electron transport occurs. The temperature dependence of electric current has also been determined in the range from 2K to 300K. Mechanical motion of the shuttle leads to the exponential temperature dependence while the temperature dependence of the electron distribution function in the leads brings about the factor $1/T$ at high temperature. The competition of these two factors gives rise to wide variety of current-temperature curves depending on quantum parameter $\nu_0$, such as decreasing of current with temperature increasing (at small $\nu_0$), a weak temperature dependence (at the intermediate values), and a strong exponential growth (at large $\nu_0$). This variety was previously demonstrated in the experiments on electron transport through long molecules \cite{Reed2,Stewart1}.

\begin{figure}[tbp]
\caption{Current-voltage characteristics of quantum shuttle system at $E_0 - \mu = 0.2 \omega_0$ and at various temperatures; (a) for quantum parameter $\nu_0=0.7$,
(b) for quantum parameter $\nu_0=0.07$. Inset: Instability of electron transport at higher voltages for various separations of the electron energetic level in the shuttle and equilibrium chemical potential of the leads.}
\label{fig1}
\end{figure}

\begin{figure}[tbp]
\caption{Temperature dependence of the electric current through the shuttle for various values of quantum parameter $\nu_0$.} 
\label{fig2}
\end{figure}


\begin{references}
\bibitem{eVd} We have neglected a term $eVx/d$ which only leads to small shift of the oscillator position, $(\delta x)_V\ll (d/2) ,$ if $V \ll m\omega_0^2d^2/2e.$ Here, $d$ is a distance between leads. 
For the system considered in the present paper with $m=10^{-20}g,\omega_0/2\pi = 1 THz,d = 3 nm,$ this voltage should be much less than $10^4 V.$
\end{references}
\end{document}